\documentclass{aa}

\usepackage{txfonts}
\usepackage{graphicx}

\begin{document}

\title{Swift observations of the prompt X-ray emission and afterglow from
  GRB050126 \& GRB050219A.}


\author{M. R. Goad\inst{1} \and G. Tagliaferri\inst{2} \and K.L. Page\inst{1}
  \and  A. Moretti\inst{2} \and J.P. Osborne\inst{1} \and S. Kobayashi\inst{3}
  \and P. Kumar\inst{3} \and P.I. M{\'e}sz{\'a}ros\inst{3} \and
  G. Chincarini\inst{2,6} \and T. Sakamoto\inst{4} \and B. Zhang\inst{5} \and
  S.D. Barthelmy\inst{4} \and A.P. Beardmore\inst{1} \and D.N. Burrows\inst{3}
  \and S. Campana\inst{2} \and M. Capalbi\inst{7} \and L. Cominsky\inst{4} \and G. Cusumano\inst{8} \and N. Gehrels\inst{4} \and P. Giommi\inst{7} 
\and O. Godet\inst{1} \and J.E. Hill\inst{3,4,9} \and J.A. Kennea\inst{3} \and
  H. Krimm\inst{4} \and V. La Parola\inst{8} \and V. Mangano\inst{8} \and
  T. Mineo\inst{8} \and D.C. Morris\inst{3} \and K. Mukerjee\inst{1} \and
  J.A. Nousek\inst{4} \and P.T. O'Brien\inst{1} \and C. Pagani\inst{2,3} \and
  M. Perri\inst{7} \and P. Romano\inst{2} \and A. A. Wells\inst{1} }

\institute{Department of Physics and Astronomy, University of Leicester,
LE1 7RH, UK 
\and INAF-Osservatorio Astronomica di Brera, via Bianchi 46, 23807
Merate, Italy \and Department of Astronomy \& Astrophysics, 525 Davey Lab., 
Pennsylvania State University, University Park, PA 16802, USA
\and NASA Goddard Space Flight Center, Greenbelt, MD 20771, USA
\and Department of Physics,
University of Nevada, BOX 454002, Las Vegas, NV 891, USA
\and Universita degli studi di Milano-Bicocca, P.za dell Scienze 3, I-20126 Milano, Italy
\and ASI Science Data Center, Via Galileo Galilei, I-00044 Frascati,
Italy.
\and INAF - Instituto di Astrofisica Spaziale e Cosmica, Via Ugo La
Malfa 153, I-90146, Palermo, Italy
\and Universities Space Research Association, 10211 Wincopin Circle,
Suite 500, Columbia, MD, 21044-3432, USA }

\date{Received : 02/11/2005/ Accepted : 23/11/2005 }

\abstract{We report on the temporal and spectral characteristics of the early
X-ray emission from the Gamma Ray Bursts GRB050126 and GRB050219A as observed
by {\it Swift\/}. The X-ray light-curves of these 2 bursts both show
remarkably steep early decays ($F(t)\propto t^{-3}$), breaking to flatter
slopes on timescales of a few hundred seconds. For GRB050126 the burst shows
no evidence of spectral evolution in the 20-150 keV band, and the spectral
index of the $\gamma$-ray and X-ray afterglows are significantly different
suggesting a separate origin. By contrast the BAT spectrum of GRB050219A
displays significant spectral evolution, becoming softer at later times, with
$\Gamma$ evolving toward the XRT photon index seen in the early X-ray
afterglow phase.  For both bursts, the 0.2-10~keV spectral index pre- and
post-break in the X-ray decay light-curve are consistent with no spectral
evolution.  We suggest that the steep early decline in the X-ray decay
light-curve is either the curvature tail of the prompt emission; X-ray flaring
activity; or external forward shock emission from a jet with high density
regions of small angular size ($>\Gamma^{-1}$). The late slope we associate
with the forward external shock.

\keywords{gamma-ray: bursts -- Gamma-rays, X-rays: individual(GRB050126, GRB050219A)}}

\titlerunning{Swift observations of GRB050126 and GRB050219A}
\authorrunning{Goad et al.}
\maketitle

\section{Introduction}

The {\it Swift\/} Gamma-Ray Burst Explorer (Gehrels et~al. 2004), launched
2004 November 20 is now routinely observing the prompt gamma-ray and afterglow
emission of Gamma-Ray Bursts (GRBs) in the astrophysically important minutes
to hours timescale after the burst trigger. With its unique autonomous
pointing capability, {\it Swift\/} is able to slew its narrow-field
instruments, the X-Ray Telescope (XRT, Burrows 2005) and UltraViolet-Optical
Telescope (UVOT, Roming et~al. 2005) to the burst position on timescales of
less than 100~s, opening up to scrutiny a largely unexplored region of
parameter space.

The greater sensitivity over previous gamma-ray missions of the Burst Alert
Telescope (hereafter BAT, Barthelmy 2004) together with {\it Swift\/}'s rapid
pointing capability, allows prompt localisations ($\sim$ a few arcseconds) of
relatively faint GRB afterglows, essential for ground-based follow-up. {\it
Swift\/} can therefore not only study fainter bursts, but also the early
afterglows during a phase in which they are many orders of magnitude brighter.
During the course of its 3-year mission {\it Swift\/} will deliver unique
insights as to the nature of the prompt and early afterglow emission as well
as probing the faint end of the GRB luminosity function.

Here we report on {\it Swift\/} observations of GRB050126 and GRB050219A (see
also Tagliaferri et~al. 2005; Nousek et~al 2005; and Chincarini et~al. 2005,
for a discussion of the average early X-ray light-curve behaviour of a sample
of Swift detected GRBs), two bursts for which we have early observations
(within $\sim$100~s) in the XRT and which show similar behaviour in their
prompt X-ray light-curves.  In \S2 we describe the data taken for each burst
in succession and present in detail a temporal and spectral analysis of the
BAT and XRT data.  In \S3 we place the observations in the context of
theoretical models of the GRB and afterglow emission. Our conclusions are
presented in \S4.

\section{Observations}

\subsection{GRB050126}

The $Swift$ BAT triggered on GRB050126 at 12:00:54 UT Jan 26th 2005 (Sato
et~al. 2005). The spacecraft autonomously slewed to the burst location and was
settled on target at 12:03:04 UT.  Before the slew, the XRT was in manual
state taking calibration observations of Mkn~876 in Photon Counting (PC) mode.
Thus the first 278~s of observations were taken in PC mode only. Ground
analysis of the early PC mode data identified a new bright source at position
RA(J2000) 18:32:27.0, Dec(J2000) +42:22:13.5 with an estimated uncertainty of
8 arc-seconds (Kennea et~al. 2005; Campana et~al. 2005a,b).

\subsubsection{BAT spectrum and light-curve of GRB050126}

Analysis of BAT calibration targets during the mission verification phase
shows that the early BAT response matrix (build 11) under-predicts the flux by
20-30\% at energies below 20~keV and predicts an excess of emission at
energies above 100~keV.  The BAT event data described here were re-analysed
using the standard BAT analysis software (build 14) as described in the Swift
BAT Ground Analysis Software Manual (Krimm, Parsons, \& Markwardt 2004), which
incorporates post-launch updates to the BAT spectral response and effective
area and includes the systematic error vector which must be applied to all
BAT spectra\footnote{see
http://swift.gsfc.nasa.gov/docs/swift/analysis/bat\_digest.html}.

\begin{figure}
\resizebox{\hsize}{!}{\includegraphics[angle=270,width=8cm]{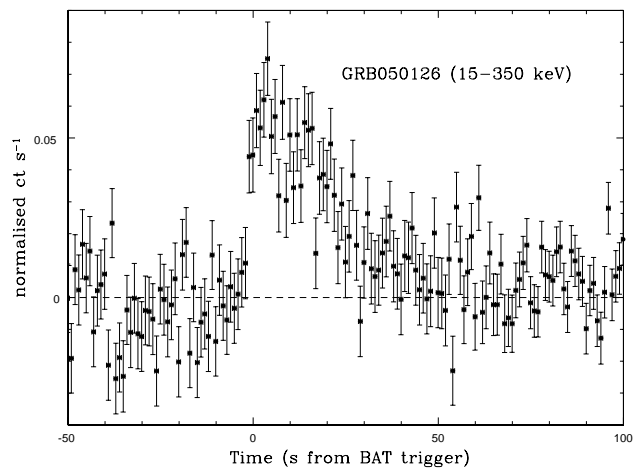}}
\caption{15-350~keV mask-weighted (ie. background subtracted) BAT lightcurve of
GRB050126. The x-axis indicates the elapsed time in seconds from the BAT rate trigger.
\label{GRB050126_batlc}}
\end{figure}

GRB050126 is characterised by a single broad peak with fast rise ($\sim1$~s),
and exponential decay (FRED) and total duration $\sim 30$~s
($T_{90}=25.7\pm0.1$~s; Fig~\ref{GRB050126_batlc}).  The total fluence is
($1.7\pm0.3)\times10^{-6}$~erg~cm$^{-2}$ (15--350~keV) and the peak five
second flux is 0.4~ph~cm$^{-2}$~s$^{-1}$.  The spectrum is
consistent with an unabsorbed power-law with photon index $\Gamma=1.44\pm0.18$
(where $P(E)\propto E^{-\Gamma}$) in the 20--150 keV band, $\chi^{2}=66.2$ for
53 dof (Sato et~al. 2005). A cut-off powerlaw does not significantly
improve the fit ($\chi^{2}=66.4$ for 52 dof), and the high-energy cut-off is
unconstrained by the data (Table~\ref{table_bat}). We find no evidence for
spectral evolution in the BAT data for this source. From 12:02:32 UT onward
(ie. 98 s after the BAT trigger), there is no detectable gamma-ray emission
with an approximate $2\sigma$ upper limit of
$7\times10^{-9}$~erg~cm$^{-2}$~s$^{-1}$ (15--350 keV).

\subsubsection{XRT observations of GRB050126}

Because the XRT was in Manual State before the slew, the standard set of XRT
observations was not implemented and thus the image mode (IM) observations
normally taken once the spacecraft has settled were not taken in this
instance. Furthermore, in Manual State automatic mode-switching is disabled
hence only photon counting (PC) mode observations were taken for this source.  

The PC mode event lists were processed using the standard {\bf xrtpipeline}
data reduction software, version 12, within FTOOLS v5.3.1, screening for
hot-pixels, bad columns and selecting event grades 0-12 for light-curves and
grade 0 for spectra.  Early calibration observations show that XRT can suffer
from a high optical background light that dominates the spectrum at low
energies ($<0.2$~keV), and which is particularly strong near the bright Earth
limb. In the analysis presented here, the low energy photons have been removed
by filtering out PHA values below 0.2~keV. Bad columns, and channels above
10~keV have also been excluded.  Finally, frames with spacecraft pointing
directions greater than 4.8 arcminutes from the nominal GRB position, and with
CCD temperatures $>-$50 degrees Celsius have also been
excluded.\footnote{Failure of the Thermo Electric Cooler (TEC) during the
verification phase of the mission has resulted in operational temperatures for
the XRT CCD in the range $-40$ to $-70$ degrees Celsius.  While above the
nominal operational temperature of $-100$ degrees Celsius, the XRT operates
well within pre-flight specifications for temperatures below $-50$ degrees
Celsius.} For light-curve and spectral extraction we define both an annular
source region of inner radius 3 pixels and outer radius 30 pixels to account
for the moderate pile-up at early times, and a circular source region of
radius 30 pixels ($\equiv71$'') at later times, both centred on the XRT
position as determined from the XRT analysis task {\bf xrtcentroid}. For the
background we define an annular region of inner radius 80 pixels and outer
radius 120 pixels centred on the same position.  Our selection criteria
yielded a total on-source time of 8076~s from 9 orbits of PC mode data, with a
mean background count-rate of 0.0019~ct~s$^{-1}$.

\begin{figure}
\resizebox{\hsize}{!}{\includegraphics[angle=270,width=8cm]{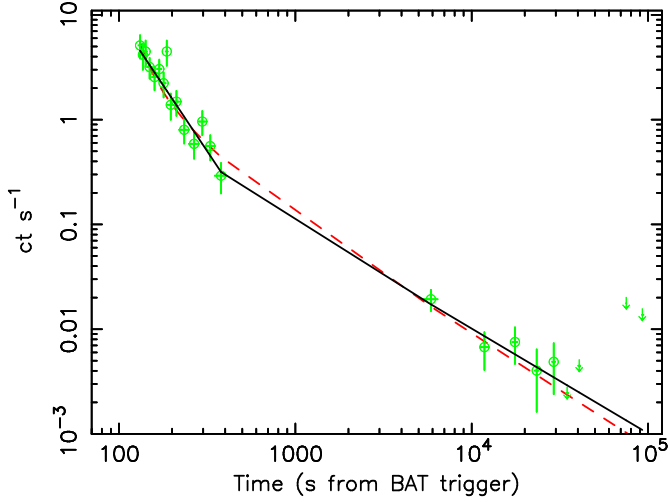}}
\caption{The XRT
  0.2--10.0~keV decay light-curve for GRB050126. All data were taken in PC
  mode.  The decay curve is well-fit by a broken power-law (solid line) with
  slope $2.52^{+0.50}_{-0.22}$ at early times flattening to a slope of $1.0$
  after $424$~s (Cash statistic$ = 26.1$ for 20 data points). Also shown is a
  powerlaw fit to the decay light-curve accounting for the onset of the X-ray
  afterglow $t_{\rm a}$ (dashed line), ie. $f_{\nu}\propto(t-t_{\rm
    a})^{-\alpha}$, with a best-fit slope of $\alpha=1.08^{+0.09}_{-0.09}$ and
  $t_{a}=105.1^{+9.1}_{-11.3}$~s and Cash statistic 31.7 for 20 data
  points. The green arrows at later times are $1\sigma$ upper limits.}\label{GRB050126_lc}
\end{figure}

\subsubsection{XRT light-curve and spectra of GRB050126}

Fig~\ref{GRB050126_lc} shows the XRT light-curve for GRB050126. 
For the first 2 orbits of data, the light-curve points were grouped
to a minimum of 20 counts/bin, while, at later times, we bin data into a 
single bin for each orbit. The source and background lightcurves were then
simultaneously fitted within XSPEC, using Cash Statistics since many of the
latter data points do not contain the minimum number of counts required for
Gaussian statistics.

Here we chose to use XSPEC for light-curve fitting as it provides a ready-made
suite of models with which to fit the data.\footnote{XSPEC does not bin
data. It groups data together between start and stop times such that a
specified number of counts are contained within that bin. XSPEC then
integrates a model fit to the data. The difference between an XSPEC fit and a
fit to binned data is rather subtle, the main difference being that the XSPEC
fit to the data is not constrained to pass through the centre of the time bin.
This difference is extremely important when fitting GRB light-curves, as they
decay rather rapidly, typically $f_{\nu}(t)\propto t^{-1}$ or faster. The use
of overly large bins at early times can drastically alter the derived slope of
the light-curve if model integration is not performed.}.
\begin{figure}
\resizebox{\hsize}{!}{\includegraphics[angle=270,width=8cm]{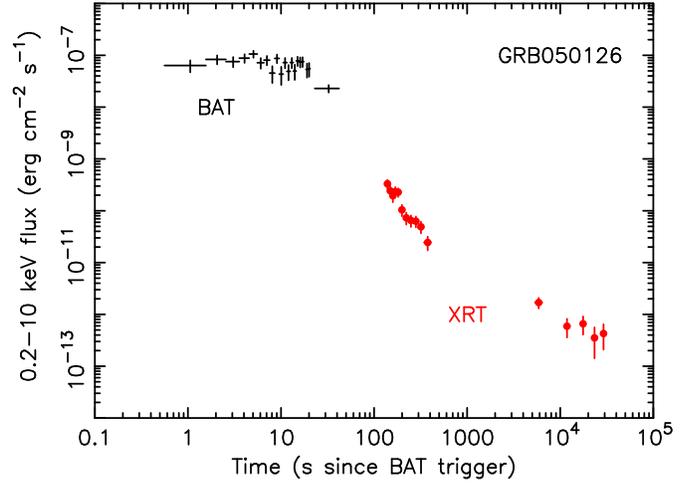}}
\caption{BAT-XRT decay light-curve for GRB050126 (0.2--10~keV, see text for
  details). Note that the early steep decline phase in the XRT light-curve 
joins smoothly with the late-time BAT data.\label{GRB050126_BAT_XRT_spec}}
\end{figure}
The 0.2--10.0~keV XRT light-curve of GRB050126 shows a steep early decline
breaking to a flatter decay slope on relatively short timescales (a few
hundred seconds). A single power-law fit to the XRT light-curve
($f_{\nu}(t)\propto t^{-\alpha}$), with $\alpha=2.5^{+7.5}_{-0.3}$, Cash
statistic = 62.0 for 20 data points, provides a poor fit to the data.  A
broken power-law provides an improved fit to the data (Cash statistic = 26.1
for 20 data points) yielding an initial slope $\alpha=2.52^{+0.50}_{-0.22}$,
flattening to $\alpha=1.00^{+0.17}_{-0.26}$ after a time
$T_{\rm{b}}+424^{+561}_{-120}$~s (Table~\ref{table_decay}, column 1),
where $T_{\rm{b}}$ is the rate trigger time and all errors are 90\%
confidence for 1 interesting parameter. We note that because of a gap in the
data coverage due to an early SAA passage, only the lower limit to the break
timescale is well-constrained. A single powerlaw fit to the data which allows
for differences between the burst rate trigger and the onset of afterglow
$t_{a}$, (ie. $f_{\nu}(t)\propto (t-t_{a})^{-\alpha}$), provides an acceptable
fit to the data with $\alpha=1.08^{+0.09}_{-0.09}$,
$t_{a}=105.1^{+9.1}_{-11.3}$, Cash statistic = 31.7 for 20 data points.
 
Ancillary response files were created using the XRT analysis task {\bf
xrtmkarf} (version 12).  The XRT spectrum is well fit by an
absorbed\footnote{Here we use the XSPEC {\bf wabs\/} Wisconsin absorber model}
power-law, photon index $\Gamma=2.26^{+0.26}_{-0.25}$, with the column fixed
at the Galactic value of $N_{\rm H}\sim 5.3\times 10^{20}$~cm$^{-2}$, yielding
$\chi^{2}=8.2$ for 8 degrees of freedom. An excess column above the Galactic
value is not required by the data. We have also formed time-averaged spectra
before and after the measured break in the XRT decay light-curve
(Table~\ref{table_xrt}).  With the column fixed at the Galactic value, and
photon index $\Gamma$ pre- and post-break tied together we find a marginally
steeper (though consistent within the errors) mean photon index of
$\Gamma=2.42^{+0.33}_{-0.31}$, $\chi^{2}=8.6$ for 6 degrees of
freedom. Untying $\Gamma$ pre- and post-break, we find marginal evidence (90\%
confidence) for spectral hardening in the 0.2--10.0~keV band following the
break in the decay light-curve, with $\Gamma$ decreasing from
$2.59^{+0.38}_{-0.35}$ pre-break to $1.72^{+0.65}_{-0.60}$ post-break.
Furthermore, by fixing $\Gamma$ to the post-break value of $1.72$, a fit to
the pre-break data reveals an additional soft excess, which may be modelled as
a blackbody with temperature $kT\sim0.11^{+0.04}_{-0.03}$~keV, $\chi^{2}=0.5$
for 3 dof, or a powerlaw with photon index $\Gamma=3.8^{+1.9}_{-1.3}$,
$\chi^{2}=1.1$ for 3 dof. We emphasise that while indicative of a possible
excess in emission at low energies, the quality of the statistics is poor (only
3 degrees of freedom), therefore we prefer not to draw any firm conclusions
based on this finding.
\begin{figure}
\resizebox{\hsize}{!}{\includegraphics[angle=270,width=8cm]{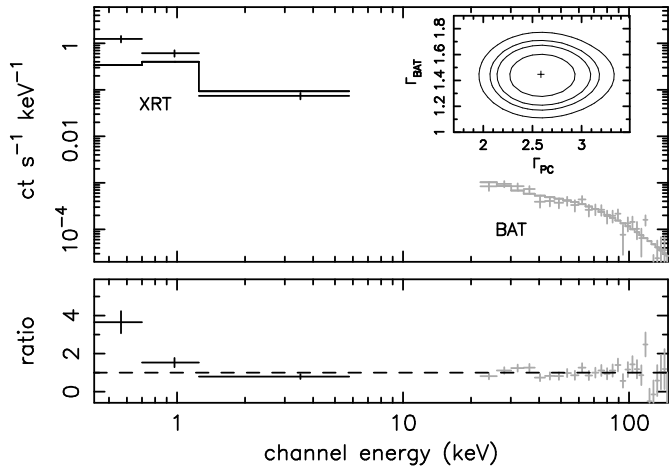}}
\caption{Combined power-law fit to the BAT (20--150 keV) and pre-break XRT
  (0.2--10 keV) spectra of GRB050126. The data are inconsistent with a single
  powerlaw fit at greater than $>$99\% confidence. Inset : 68\%, 90\%, 95\%,
  and 99\% confidence contours indicating the Photon indices for BAT and XRT
  PC mode data in a combined fit to the BAT and XRT spectra for GRB050126.
  The data are inconsistent with a single power-law fit to the BAT and XRT
  spectra at $>$99\% confidence.\label{GRB050126_bat_xrt_spec}}
\end{figure}

In Fig~\ref{GRB050126_BAT_XRT_spec} we show the combined BAT and XRT flux
light-curves corrected for absorption. The BAT photon index is insensitive to
the Galactic column for energies above 20~keV. Therefore to derive the BAT
flux light-curve we have simply extrapolated the BAT spectrum into the XRT
band (0.2--10~keV) adopting the average photon index of the pre-slew BAT and
pre-break XRT spectra ($\Gamma=2.02$).

Taken at face value, the pre-break XRT light-curve appears to point toward the
late-time BAT data indicating that the early X-ray emission may be associated
with the end of the initial explosion.  A combined powerlaw fit to the BAT
(20-150~keV) and pre-break XRT (0.2-10~keV) spectra
(Fig~\ref{GRB050126_bat_xrt_spec}), covering the time interval $T_{\rm
b}-0.3$~s to $T_{\rm b}+29.5$~s for the BAT spectrum and $T_{\rm b}+130$~s to
$T_{\rm b}+250$~s for the XRT spectrum, however, is inconsistent with a single
powerlaw slope at $>$99\% confidence, with the XRT spectrum being
significantly softer. This does not preclude an association between the prompt
and early X-ray emission since these observations are non-simultaneous, and
the prompt emission may evolve at later times, though we note that there is no
evidence for spectral evolution in the BAT data.

\subsubsection{Observations of GRB050126 in other bands}

Due to the close proximity of GRB050126 to the bright star Vega, UVOT was
unable to observe this field.  Ground based observations with Keck/NIRC
obtained on Jan 26th, 4.5 hrs after the burst (Berger \& Gonzalez 2005)
revealed a new infra-red (Ks band) source 1.9 arcseconds from the XRT position
with a subsequent redshift determination for the host galaxy of $z=1.29$.


\subsection{GRB050219A}

The {\it Swift\/} BAT triggered on GRB050219A at 12:40:01 UT Feb 19th 2005
(Hullinger et~al. 2005). The spacecraft autonomously slewed to the burst
location starting at $T_{b}+12$~s and was on target at $T_{b}+78$~s. Following
the slew {\it Swift\/} began an automated sequence of observations with
XRT and UVOT.
\begin{figure}
\resizebox{\hsize}{!}{\includegraphics[angle=270,width=8cm]{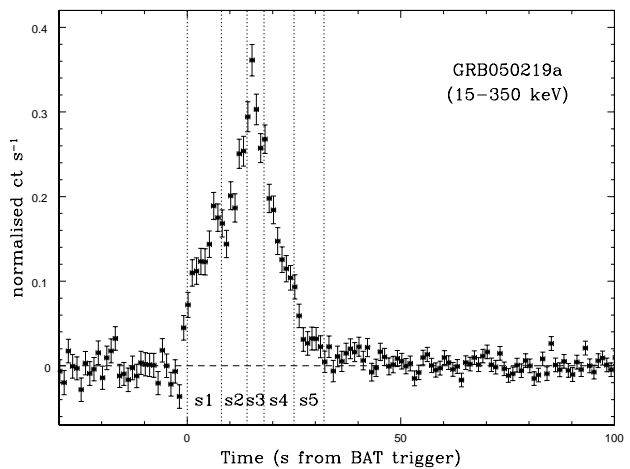}}
\caption{
15--350~keV mask-weighted BAT lightcurve of GRB050219a. The dotted lines
(segments 1-5) indicate the time-slices used for the BAT spectral evolution
analysis. Zero represents the BAT rate trigger time.\label{GRB050219a_batlc}}
\end{figure}

\subsubsection{BAT spectrum and light-curve of GRB050219A}

The BAT light-curve for GRB050219A is characterised by two overlapping peaks,
with a duration $T_{\rm 90}=23.5\pm0.02$~s (Fig~\ref{GRB050219a_batlc}).  The
peak flux is $5.5$~ph~cm$^{-2}$~s$^{-1}$ for a 1-s interval ($15-350$~keV),
15~seconds after the onset of the burst. The burst fluence is
$\sim(5.2\pm0.4)\times10^{-6}$~erg~cm$^{-2}$ in the $15-350$~keV band.  In the
20--150~keV energy range, the BAT spectrum is well-fit by a cut-off power-law
with photon index $\Gamma=-0.39^{+0.38}_{-0.40}$, high energy cut-off
$40.61^{+12.5}_{-8.2}$~keV, $\chi^{2} = 46.9$ for 52 dof. A single power-law
fit to the 20--150~keV spectrum is significantly worse ($\chi^{2}=110.2$ for
53 dof, Table~\ref{table_bat}).  A Band model (Band et~al. 1993) fit to the
data is poorly constrained due to the limited range and low BAT effective area
at high energies.

\begin{figure}
\resizebox{\hsize}{!}{\includegraphics[angle=270,width=8cm]{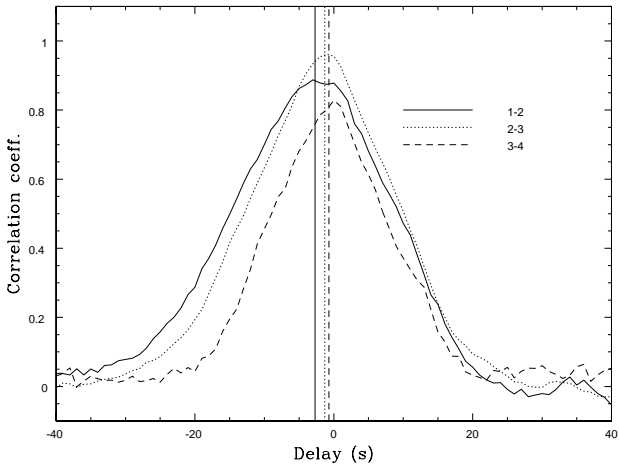}}
\caption{Cross-correlation functions indicating the interband delays in 4 BAT
channels for GRB050219a. The light-curves are highly correlated with the
hardest bands leading the softer bands by $0.7\pm0.1$~s (band 4 -- band 3),
$1.3\pm0.1$~s (band 3 -- band 2), and $2.7\pm0.1$~s (band 2 -- band 1)
respectively.}\label{xcorr}
\end{figure}

A cross-correlation analysis of the 4 channel, 1~s, BAT light-curves
(15-25~keV, 25-50~keV, 50-100~keV, 100-350~keV) shows that the 4-channel
light-curves are highly correlated, and display evidence for inter-band
delays, with the lowest energy bands delayed with respect to the higher energy
bands (Fig~\ref{xcorr}). This is confirmed by a simple hardness ratio plot,
which shows the BAT spectrum is indeed significantly softer at later times.
The highest energy band 100-350~keV precedes the 50-100~keV band by
$0.7\pm0.1$~s, the 50-100~keV precedes the 25-50~keV band by $1.3\pm0.1$~s and
the 25-50~keV band precedes the 15-25~keV band by $2.7\pm0.1$~s, all
measurements determined from a Gaussian fit to the peak (correlation
coefficient $>0.5$) of the cross-correlation function.

\begin{figure}
\resizebox{\hsize}{!}{\includegraphics[angle=270,width=8cm]{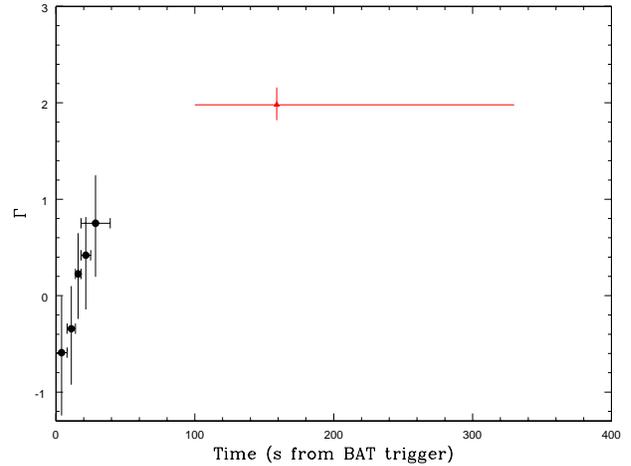}}
\caption{Evolution of the BAT photon index 15-150~keV (black points)
for GRB050219a during segments 1-5 (see text for details).  The red triangle
indicates the 0.2--10~keV XRT photon index during the steep early decline
phase of the burst. The BAT and XRT errors are 90\% confidence limits on the
spectral fit.}\label{bat_xrt_evol}
\end{figure}

To determine the rate at which the photon index evolves, we have divided the
15-350~keV light-curve into 5 segments, fitting the spectrum in each segment
with a cut-off powerlaw (Table~\ref{table_grb050219a}).  A single cut-off
powerlaw fit to all 5 segments is not a good fit to the data, with
$\Gamma=0.25$, $E_{\rm peak}=82.8$~keV, $\chi^{2}=821$ for 272 dof.  Keeping
the photon index tied between each segment and allowing the cutoff energy to
freely vary, provides a significant improvement, with $\Gamma=0.012$, and
$E_{\rm peak}$ increasing from 77.5~keV to 139.2~keV through segments 1-3
before decreasing in the final 2 segments to 59.6~keV, $\Delta \chi^{2}=539$
for 268 dof. Finally, allowing the photon index to freely vary provides
further improvement, $\Delta \chi^{2}=45$.  In Table~4 we show cut-off
powerlaw fits to individual segments only.  Table~4 confirms our earlier
finding that the BAT photon index steepens at later times. As with the
combined fit we find that the peak energy in the cut-off powerlaw appears to
increase through segments 1-3, peaking in segment 3, when the burst was
brightest, before shifting back to lower energies in the final segment. The
observed decrease in peak energy at later times is a feature common to many
burst afterglows (Ford et~al. 1995, Norris et~al. 1986).
Fig~\ref{bat_xrt_evol} shows the evolution of the BAT photon index
(15-350~keV) with time together with the measured photon index in the XRT
0.2-10~keV during the early steep decline phase of the burst (see \S\ref{xrt}
and \S2.2.3).

\subsubsection{XRT observations of GRB050219A}\label{xrt}

\begin{figure}
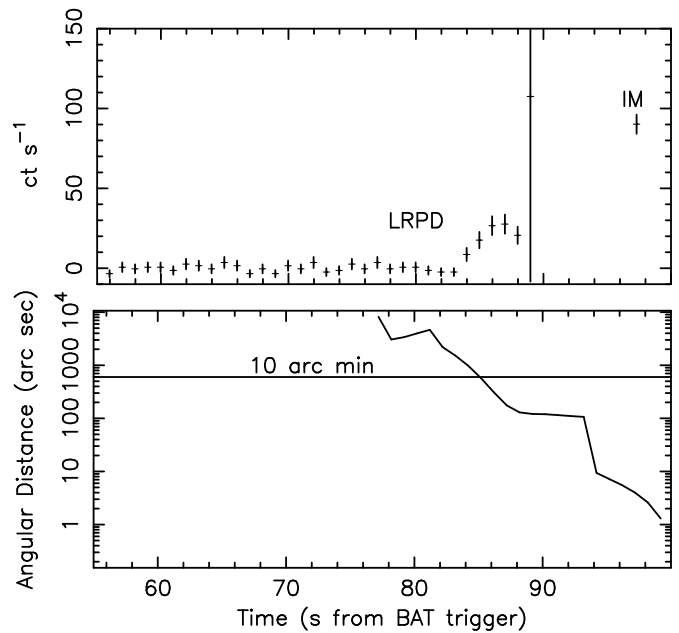

\resizebox{\hsize}{!}{\includegraphics[angle=270,width=7cm]{4457fig8a.eps}}
\resizebox{\hsize}{!}{\includegraphics[angle=270,width=7cm]{4457fig8b.eps}}
\vspace{5mm}
\caption{Timeline showing the LrPD count-rate during the slew, settling and
early pointing phase observations of GRB050219a. The object enters the XRT
field of view $\approx~82$ seconds after the BAT trigger. Observations at
$t_{\rm b}+86$ s onwards are less than 5 arcminutes from the final pointing
and are well within the XRT field of view ($24\times24$ arcminutes
squared).\label{settling}}
\end{figure}

The XRT was in Auto-State when {\it Swift\/} slewed to the burst.  During the
slew XRT was in Low rate PhotoDiode (LrPD) mode and the early frames are
vignetted (Fig~\ref{settling}). After settling, a single exposure (2.5~s
duration) Image Mode frame was taken, truncating the final LrPD mode
frame. XRT then cycled through Windowed Timing (WT) mode and PC mode
observations, with the mode determined by the count-rate and the on-board
switch thresholds (Hill et~al. 2005). The first usable data, an un-vignetted
LrPD frame, was taken 86.5~s after the burst trigger.  The LrPD data indicate
an initial increase in the observed count rate over and above that of the
background count rate, which we assume to be due to the source moving into the
field of view of the XRT (see Figure~8). GRB050219A was observed for a total
of $\sim 3730$ seconds.  XRT returned to the target in the following orbit for
a second pointed observation in PC mode (total duration 2170~s). No further
observations of GRB050219A were taken during the next few weeks as GRB050219b
(Cummings et~al. 2005) became the new priority target. Observations taken in
March did not detect the source.

The event lists for the various XRT observing modes were processed using the
standard xrtpipeline calibration software, version 12, screening for
hot-pixels, bad columns and using grade selection: PC mode 0-12,
WT mode 0-2, LrPD mode 0-5.  We note that for the first $\sim 1000$~s the CCD
temperature was marginally above $-50$~Celsius. However, as far as we can
tell, there do not appear to be any temperature-dependent artefacts in the
data.

\noindent{\it LrPD mode data}

In Low rate PhotoDiode mode the charge from a full CCD frame is accumulated
into a single pixel and thus all spatial information is lost. The timing
resolution is however excellent ($0.14$~ms). A total of four 8.3~s LrPD
mode frames were taken during the slew with approximately 3 seconds of usable
data (ie. data for which the angular distance $< 5$ arcminutes, see
Fig~\ref{settling}).  LrPD light-curves have been extracted using the standard
grades (0-5) and selecting energies between 0.2--10.0~keV.  After binning to
1~s bins, we determine a mean background count-rate in LrPD mode using the
first 25 seconds of data. The mean background rate is then subtracted from all
LrPD mode data.  The final LrPD frame is truncated due to the switch to Image
Mode and consequently the final 1~s bin has a low fractional exposure.
However, the large measured count rate and associated error
($105\pm117$~ct~s$^{-1}$) are consistent with the count rates determined from
the following IM and WT mode observations extrapolated back to the
time of the LrPD frames. After settling, one additional 8.3~s LrPD frame was
taken before switching to WT mode.

\noindent{\it Image Mode data}

Using simple aperture photometry on the Image Mode frame we determined the
integrated DN above the background in a 30-pixel wide circular aperture
centred on the source. For GRB050219A there were 7795 DN above the background
rate (136~DN) in the 2.5~s exposure. DN are converted to count rates using the
spectral fit to the WT mode data to estimate the average energy of a count in
the spectrum in the 0.2--10~keV band ($2731$~eV). For low gain image mode
data, the mean energy per DN is $\sim79$~eV, thus giving 34 DN/ct. The total
number of counts per second above the background is therefore
$90.2\pm6$~ct~s$^{-1}$. Scaling the counts to the mean flux in the 0.2-10~keV
band gives a total observed flux in the Image Mode frame of
$5.36\pm0.36\times10^{-9}$~erg~cm$^{-2}$~s$^{-1}$.

\noindent{\it PC and WT mode data}

For source extraction we define two extraction windows, one for the source and
one for the background. For source extraction in PC mode we define a 30 pixel
radius ($\equiv71$'') circular region centred on the source position as
measured using the task {\bf xrtcentroid}, and an annular region, inner radius
30 pixels, outer radius 50 pixels, centred on the same position for the
background. For WT mode we use two rectangular regions each 40 pixels long,
the first centred on the source position, the other located in a separate
background region. Ancillary response files were created using the Swift
software data analysis task {\bf xrtmkarf}.

\subsubsection{XRT light-curve and spectra of GRB050219A}

\begin{figure}
\resizebox{\hsize}{!}{\includegraphics[angle=270,width=8cm]{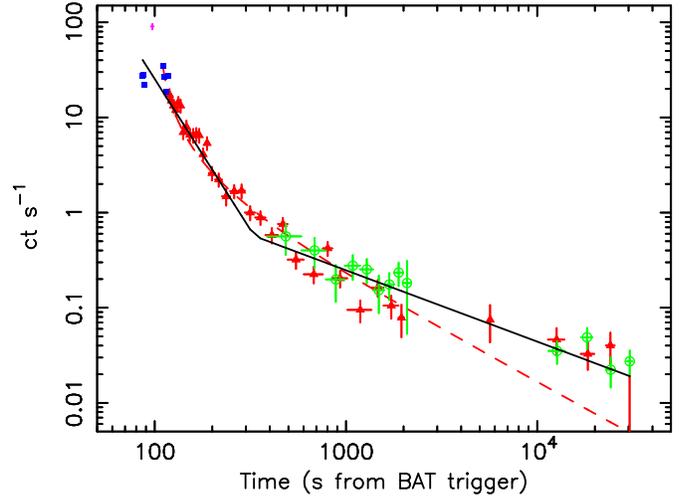}}
\caption{The XRT 0.2--10.0~keV decay light-curve for GRB050219a. Colours
 indicate LrPD data (blue), IM data (magenta), WT data (red), and PC mode data
 (green). The best-fit broken power-law fit to the decay curve after 100
 seconds has a decay slope $\alpha_{1}=3.17^{+0.24}_{-0.16}$ at early times
 flattening to a slope of $\alpha_{2}=0.75^{+0.09}_{-0.07}$ after
 $332^{+26}_{-22}$~s (solid line), $\chi^{2}=74.6$ for 36 dof. We also show
 (dashed line) the result of fitting the function $f_{\nu}(t)\propto (t-t_{\rm
 a})^{-\alpha}$, where $t_{a}$ represents the onset of the afterglow, with a
 best-fit slope $\alpha=1.10^{+0.09}_{-0.08}$ and
 $t_{a}=100.7^{+2.8}_{-4.0}$~s. This model is a poor approximation to the late
 time data ($t>5000$~s). We note that the early LrPD and IM data suggest a
 possible flare at early times, or alternatively, may indicate the onset of
 the afterglow emission.\label{GRB050219a_decay}}
\end{figure}

Figure~\ref{GRB050219a_decay} shows the XRT light-curve constructed from the
standard grade selections for all modes, and grouped for a minimum of 40
counts/bin. Note we do not bin across gaps in the light-curves caused by
orbital viewing constraints, nor do we include data taken within different
modes within the same bin.

The prompt X-ray emission, detected within 86~s of the BAT rate trigger, is
one of the earliest X-ray detections of a GRB made by {\it Swift\/}.  If we
include the LrPD data taken during the slew, the XRT lightcurve indicates an
initial rise, possibly due to an X-ray flare, followed by a steep decline from
a peak near the time of the Image Mode exposure. Flaring behaviour has been
detected in approximately half of the Swift GRB population (see e.g. Burrows
et~al. 2005; Barthelmy et~al. 2005).  For the decline phase, we measure an
initial decay slope for the XRT light-curve (excluding the PC mode data)
relative to the BAT rate trigger of $\alpha_{1}=3.17^{+0.24}_{-0.16}$ at early
times, breaking to a slope of $\alpha_{2}=0.75^{+0.09}_{-0.07}$,
$332^{+26}_{-22}$~s later, with $\chi^{2}=74.6$ for 36 dof
(Fig~\ref{GRB050219a_decay}).  A single powerlaw fit to the data with the
afterglow coincident with the onset of the burst provides a poor fit to the
data with a best-fit slope of $2.50^{+0.16}_{-0.16}$, $\chi^{2} = 225$ for 38
dof.  Fitting for the onset of the afterglow, we obtain a decay slope
$\alpha=1.10^{+0.09}_{-0.08}$, $t_{a}=100.7^{+2.8}_{-4.0}$~s (90\% confidence)
after the BAT rate trigger with $\chi^{2}/dof=114.1/37$. We note that this fit
is a poor approximation to the late-time data.

\begin{figure}
\resizebox{\hsize}{!}{\includegraphics[angle=270,width=10cm]{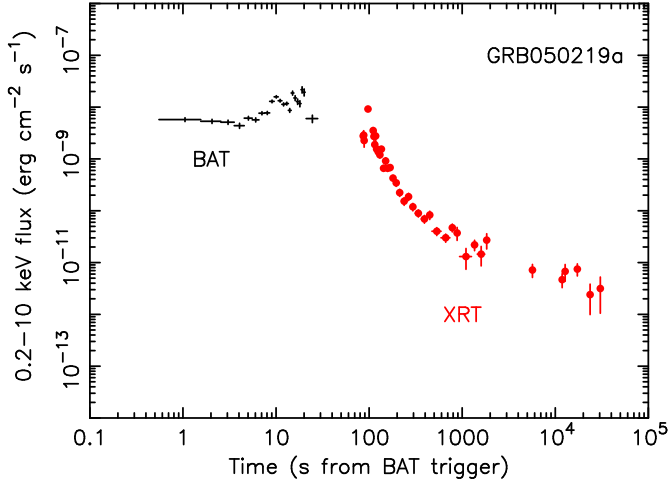}}
\caption{BAT-XRT light-curve for GRB050219a (0.2--10~keV, see text for
  details). Note that the early steep decline in the XRT data does not appear
  to join up smoothly with the late time BAT
  data.\label{GRB050219a_bat_xrt_lc}}
\end{figure}

Fitting the time-averaged spectra for the LrPD, WT and PC mode data together
yields a best-fit powerlaw model with photon index
$\Gamma=1.9^{+0.17}_{-0.16}$ with an excess absorption above the Galactic
value ($N_{\rm H}=8.5\times 10^{20}$~cm$^{-2}$) of $1.34^{+0.51}_{-0.47}\times
10^{21}$~cm$^{-2}$, $\chi^{2}=55$ for 52 dof.  We find no evidence for
spectral hardening following the break in the decay light-curve
(Table~\ref{table_xrt}) and applying the post-break spectral fit to the
pre-break data does not in this case reveal any excess emission at soft
energies. Fig~\ref{GRB050219a_bat_xrt_lc} shows the BAT and XRT decay
light-curves. Count rates have been converted to fluxes using the spectral
fits to the BAT and XRT data and assuming no spectral evolution pre- and
post-break in the XRT light-curve. The BAT data have been extrapolated into
the XRT 0.2-10.0 keV band using an average of the photon index of the BAT and
XRT data for each of the 4 segments of the BAT light-curve. For GRB050219A the
XRT light-curve does not appear to join smoothly with the late-time BAT data.
This apparent mismatch in the BAT/XRT light-curves is most likely due to
flaring behaviour in the early XRT data, although the strong spectral
evolution in the BAT data could also have some influence.

\begin{figure}
\resizebox{\hsize}{!}{\includegraphics[angle=270,width=8cm]{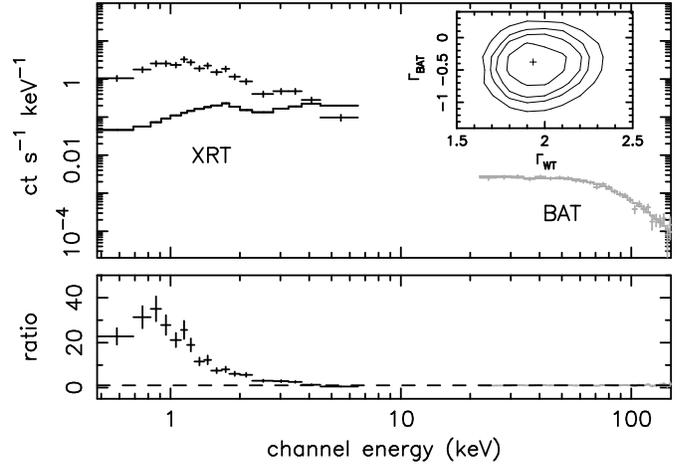}}
\caption{Combined power-law fit to the BAT (20--150 keV) and XRT (0.2--10 keV)
spectra of GRB050219a. Inset : 68.3\%, 90\%, 95.4\%and 99\% confidence contours indicating the
  relative Photon indices for BAT and XRT in a combined fit to the BAT and XRT
  spectra of GRB050219a.  The data are inconsistent with a single power-law
  fit to the BAT and XRT spectra at greater than 99\% confidence.
\label{GRB050219a_bat_xrt_spec}}
\end{figure}

Fig~\ref{GRB050219a_bat_xrt_spec} shows a combined fit to the BAT and
pre-break XRT spectra.  A single powerlaw fit to the combined BAT-XRT spectrum
is a poor ($\chi^{2}/dof = 164/103$) description of the data, again indicative
of strong spectral evolution in the GRB spectrum. We note that GRB flares
observed by Swift have generally harder canonical photon indices than spectra
derived from the underlying powerlaw decay curve and soften with time,
reminiscent of the behaviour of the prompt emission (see
e.g. Burrows et~al. 2005b).


\subsubsection{Observations of GRB050219A in other bands}
UVOT observations of GRB050219A found no new source within the XRT error
circle down to a limiting magnitude of ~20.7 in V in a combined 374~s exposure
starting 96 seconds after the BAT trigger (Schady et~al. 2005).
Ground-based observations of the XRT position for this source taken $\approx
2$~hours after the burst failed to find a counterpart down to a limiting
magnitude of $\approx20.5$ in R (de Ugarte Postigo et~al. 2005). A 20-min
observation in the I-band taken 17 hours after the burst with the 40~inch Las
Campanas Swope telescope did not detect any new source down to a limiting
magnitude of 21.5.  To date, no optical/IR or radio counterpart to this source
has been reported.

\section{Discussion}

\subsection{XRT light-curves}

The XRT decay light-curves of GRB050126 and GRB050219A are both characterised
by an initial steep decline breaking to a flatter slope on timescales of a few
hundred seconds (Table 3). A similar break, though less well-defined has also
been seen in GRB050117 (Hill et~al. 2005) and GRB050319 (Cusumano
et~al. 2005), while possibly the most convincing early break is seen in the
prompt X-ray light-curve of GRB050315 (Vaughan et~al. 2005). If these temporal
variations do indeed represent breaks in the afterglow light-curves, then they
are amongst the earliest breaks yet seen.  A single powerlaw fit to the data
with the onset of the afterglow, $t_{\rm a}$, set to the BAT trigger time, is
in both cases a poor fit to the data (Table~\ref{table_decay}).  A broken
powerlaw fit is a significant improvement in both cases, with derived break
timescales of $T_{{\rm b}}+424^{+561}_{-120}$~s (all errors 90\% confidence on
1 interesting parameter), and $T_{{\rm b}}+332^{+26}_{-22}$~s for GRB050126
and GRB050219A respectively (Table~3).  For GRB050126 we can also find an
acceptable fit to the data by fitting a single powerlaw together with an
offset ($t_{a}$) for the onset of the afterglow relative to the BAT trigger.
We find a best-fit model of $\alpha=1.08^{+0.09}_{-0.09}$,
$t_{a}=105.1^{+9.1}_{-11.3}$, Cash statistic = 31.7 for 20 data points. For
GRB050219A, a single powerlaw fit with offset $t_{\rm a}$ significantly
under-predicts the late-time (post-break) data, though the derived $t_{\rm a}$
for GRB050219A is consistent within the errors with the observed rise in the
light curve seen in the LrPD frames (Fig~\ref{settling}) taken during the
final stages of settling in the first orbit. Taking the observed rise in count
rates in the pre-slew LrPD mode frames at face value, we may well have
observed the onset of the afterglow in this burst. Since the peak count-rate
in the Image Mode frame is consistent with the sharp decline at early times,
the peak afterglow emission may have occurred prior to the Image Mode
observation.

In the previous discussion we have assumed that the early X-ray data
are entirely due to the onset of the afterglow. 
In order to quantify the relationship, if any, between the $\gamma$-ray
emission and the early X-ray light-curve, we show in Table~\ref{table_decay}
the results of fitting a Gaussian$+$powerlaw model to the XRT decay
light-curve. For both, we initially fix the late-time powerlaw decay slope to
the best-fit model for the late-time data and then fit a Gaussian assuming that
the location of the peak of the Gaussian $t_{\rm g}$ is coincident with the
burst trigger time. For both GRB050126 and GRB050219A, freeing $t_{\rm g}$
produces no improvement in the fit to the data, suggesting $t_{\rm g}$ is
indeed coincident with the BAT trigger. 

\subsection{BAT/XRT spectral fits}

The spectral indices measured for the $\gamma$-ray emission and prompt X-ray
emission immediately following the burst (i.e. during the steep decline phase)
are significantly different in both sources ($>$99\% confidence), suggesting a
separate origin. However, we note that the BAT spectral index for GRB050219A
evolves during the course of the burst toward the measured XRT photon index
(see e.g. Fig~7, and Table~2).


Fitting the XRT 0.2--10~keV band either side of the break in the light-curve
is only marginally suggestive of spectral hardening at later times,
$2.59^{+0.38}_{-0.35}\rightarrow 1.72^{+0.65}_{-0.60}$ and
$1.98^{+0.18}_{-0.16}\rightarrow 1.89^{+0.28}_{-0.23}$ for GRB050126 and
GRB050219A respectively. While we stress that in both cases the slopes pre-
and post-break are entirely consistent within the errors with no temporal
spectral evolution, we remark in passing that unlike GRB050219A, GRB050126
shows evidence of a soft excess consistent with either a thermal blackbody
with temperature $kT=0.11^{+0.04}_{-0.03}$~keV, $\chi^{2}=0.5$ for 3 dof, or a
powerlaw with $\Gamma=3.8^{+1.9}_{-1.3}$, $\chi^{2}=1.1$ for 3 dof; when
applying the spectral fit for the post-break XRT data to the pre-break XRT
data.

\subsection{Origin of the X-ray light-curve}

The X-ray afterglow emission from these two bursts consists of an early
steeply declining phase, $f_\nu(t)\propto t^{-3.0}$, lasting approximately
five minutes, followed by the more typically observed less steeply declining
phase with $f_\nu(t)\propto t^{-1}$ (see Table 2 for their observational
properties). The broken powerlaw light-curve is likely
produced by two distinct X-ray components.

The interpretation for the less steeply falling X-ray light-curve is
straightforward -- it is synchrotron radiation in the external forward
shock. The spectral index $\beta$ (where $\beta=\Gamma-1$) and the temporal
index $\alpha$ ($f_\nu\propto \nu^{-\beta} t^{-\alpha}$) are both related to
the electron powerlaw index $p$. When the observed energy band
(0.2--10~keV) is above the synchrotron cooling frequency, the most
likely possibility for these bursts at early times, then $\beta=p/2$ and
$\alpha=(3\beta-1)/2$. The values of $\alpha$ and $\beta$ for these two bursts
(see Table 5) are consistent within the errors with this expectation;
$p=3.2\pm0.7$ for GRB050126 and 1.96$\pm0.33$ for GRB050219A.

If the onset of the afterglow phase ($t_{\rm a}$) is unrelated to the burst
trigger time, then the entire X-ray afterglow light-curve for GRB050126 can be
fitted with a single power law of index $\sim1$ (Table 2) i.e., $f_\nu\propto
(t-t_{\rm a})^{-1}$ with $t_{\rm a}=105$~s (see Kobayashi et~al. 2005 for a
detailed discussion of the choice of the onset of $t_{a}$).\footnote{This is
not true for GRB050219A, for which such a model significantly underpredicts
the flux in the late-time data.} The X-ray light-curves and the spectra in
this case are entirely consistent with synchrotron radiation in the external
forward shock.  However, in the internal-external shock model $t_{\rm a}$
cannot simply be chosen arbitrarily.  According to this model the outer-most
shell, that is farthest from the central explosion, is the first to interact
with the circumstellar medium and produces the forward shock emission we see
whereas the burst itself is produced in internal shocks at smaller
radii. Therefore, we expect $t_{\rm a}$ to lie between the gamma-ray burst
trigger time and the reverse shock crossing time (afterglow peak time).  This
suggests that the break in the X-ray light-curve for GRB050219A is real and we
consider various models that can give rise to this break.

A rapidly falling X-ray light-curve at early times could be due to emisson
from a hot cocoon accompanying a relativistic jet (M{\'e}sz{\'a}ros \& Rees
2001; Ramirez-Ruiz et al.  2002), and from the photosphere associated with the
outflow from the explosion (e.g. M{\'e}sz{\'a}ros \& Rees 2000,
M{\'e}sz{\'a}ros et al. 2002, Rees \& M{\'e}sz{\'a}ros 2004). However, in the
simplest versions of these models the spectrum of the emergent radiation is
thermal which is inconsistent with the non-thermal powerlaw spectrum for the
two bursts (again we emphasise that the evidence for a soft excess with a
thermal spectrum in GRB050126 is marginal at best). Some modifications to
these models, for example, Comptonisation of the powerlaw tail of the thermal
radiation, might produce the observed behaviour.

Our preferred production mechanism for the rapidly decaying X-ray light-curves
is through internal or external shocks (see eg. Zhang et~al. 2005 for a
thorough review of the theoretical implications of X-ray afterglow
light-curves). We discuss each of these possibilities in detail below.

\subsubsection{External Shock model}

One possible explanation for the steeply falling X-ray light-curve at early
times is a high degree of angular fluctuation in the relativistic outflow
(Kumar \& Piran, 2000) or a mini-jet (Yamazaki et al. 2004).  If the angular
size of the high energy density regions, or ``bright spots", in the blast wave
is less than $\Gamma^{-1}$, the early X-ray light-curve will decline as
$t^{-p}$, as in a spreading jet case, and the spectrum is given by
$\nu^{-p/2}$ when the X-ray band is above the cooling frequency; $\Gamma$ is
the Lorentz factor of the jet.  For GRB050126 this provides a reasonable fit
to the early X-ray light-curve: $p=2\beta=3.2\pm0.7$, and
$\alpha=2.5^{+0.5}_{-0.2}$. However, for the steeply falling part of the X-ray
light-curve of GRB050219A $2\beta=1.96\pm0.16$ and $\alpha=3.17\pm0.2$, are
inconsistent with this picture (Table 5).  When the bright region along our
line of sight spreads and merges with other bright regions the subsequent
light-curve decline is the same as in the standard external forward shock
model, i.e., $\alpha\approx 1$. A possible weakness of this model is that the
probability of a random observer line of sight passing through a bright region
is smaller than the probability of that passing through the darker,
inter-bright, regions. We should therefore see many more X-ray light-curves
with $\alpha\approx 1$ at early times, or even $\alpha<0$, a conclusion which
is not supported by the data.  However, it is possible that the dark regions
are very faint in $\gamma$-rays and therefore do not trigger the
BAT.

\subsubsection{Internal or Reverse shocks}

A rapidly falling X-ray light-curve could also arise if the source activity
ends abruptly, as might be expected in the internal shock or in the reverse
shock heated GRB ejecta.  In this situation the observed flux will not drop
suddenly but will have a non-zero value for some period of time as the
observer will continue to receive radiation from those parts of the
relativistic source that lie at angles ($\theta$) greater than $\Gamma^{-1}$
with respect to the observers line-of-sight, the so-called ``curvature
effect'' (see e.g. Kumar and Panaitescu 2000; Dermer 2004; Fan and Wei
2005). For the simplest case of a uniform source the observed radiation in a
fixed observer energy band will decline with time as $t^{-2-\beta}$; the
observed spectrum is $\nu^{-\beta}$ (Kumar \& Panaitescu 2000).  The values
of $\alpha$ and $\beta$ for the steeply falling part of the X-ray light-curve
for GRB 050219A are compatible with this expectation. However, if the source
for X-rays is same as the $\gamma$-ray burst photons, which is the most
natural explanation in this model, then we expect the spectrum during the
X-ray afterglow to be the same as during the $\gamma$-ray burst.  This
requirement is violated for both of the bursts. The emission from
$\theta>\Gamma^{-1}$ from the $\gamma$-ray source provides a lower limit to
the flux we must detect at $5$ minutes in the X-ray band. This limit is also
violated for GRB050219A, suggesting that the comoving energy flux for the
$\gamma$-ray source is decreasing with increasing $\theta$.  In this case the
spectrum at $\theta>\Gamma^{-1}$ is likely to be softer than we see during the
burst, and the inconsistency between spectra during the $\gamma$-ray burst and
the X-ray afterglow could be resolved. In this scenario the rapidly falling
X-ray light-curve is from the same source as the $\gamma$-ray burst, with the
forward shock emission dominating after a few hundred seconds.

A rapidly falling X-ray light-curve is also expected from the reverse shock
heated ejecta. The synchrotron emission from the reverse shock is thought
usually to peak in the infrared or optical band (M{\'e}sz{\'a}ros \& Rees
1993, Panaitescu et al. 1998, Sari \& Piran 1999), and the optical light-curve
after the deceleration time decays as $\sim t^{-2}$. However, the optical
synchrotron photons when inverse Compton scattered in the ejecta, may emerge
in the X-ray band. The inverse Compton light-curve declines faster than the
synchrotron light-curve by a factor of $\tau_e \gamma_e^{p-1}\propto
t^{-(13p-1)/48} \sim t^{-0.5}$; where $\tau_e$ is the optical depth of the
ejecta to Thomson scattering, and $dn_e/d\gamma \propto \gamma^{-p}$ for
$\gamma>\gamma_e$.  Thus, the inverse Compton X-ray light-curve decline is
expected to be $\sim t^{-2.5}$ which is consistent with the steeply declining
early X-ray observations for these two bursts.  The optical depth of the
ejecta at deceleration is $\tau_{da}\sim 10^{-4} E_{53} n_\pm/(\Gamma^2
R^2_{16.5})$, where $n_\pm$ is the number of $e^\pm$ pairs per proton, $R$ is
the deceleration radius, and a numerical subscript $x$ means the variable
divided by 10$^x$. The optical flux at deceleration is larger than the X-ray
flux by a factor $\tau^{-1}_{da}$ and violates the upper limit on the optical
flux from these bursts unless $n_\pm > 10^2$.

If the $\gamma$-rays are generated through internal shocks, we expect the
ejecta to undergo adiabatic expansion and cool down with time once shell
collisions have ended.  As time goes by, the adiabatically expanding shells
will produce radiation that is shifted to lower and lower energies, and could
ultimately be responsible for the early rapidly declining X-ray light-curve we
observe for these two bursts. The influence of a magnetic field could enhance
this possibility.  A tangled magnetic field in the ejecta decreases with
radius as $r^{-2}$, and the electron thermal energy decreases as
$r^{-1}$. Since the Lorentz factor of the shell is not changing with time,
$r\propto t$, and the flux in a fixed observer band above the peak of the
spectrum, decays as $t^{-\alpha}$ where $\alpha=2+4\beta$, and $\beta$ is the
spectral index. For a transverse magnetic field in the ejecta, $B\propto
r^{-1}$, and the light-curve in a fixed observer band decays as
$t^{-1-3\beta}$ when the burst energy is carried by matter, or $t^{-1-2\beta}$
when the magnetic field dominates the energy in the outflow (M{\'e}sz{\'a}ros
\& Rees, 1999). We note that $\alpha$ can be no larger than $(2+\beta)$, if
the shells are homogeneous in the angular direction, because radiation from
$\theta>\Gamma^{-1}$ will always contribute to the observed flux.

For $\beta>0$, as is the case for both GRB050126 and GRB050219A, the high
$\theta$ emission dominates when the magnetic field in the shock heated ejecta
is randomly oriented. In the other two cases considered above, high $\theta$
emission dominates when $\beta\ge 1/2$ and $\beta\ge 1$ respectively. The
early X-ray spectral indices for GRB050126 and GRB050219A, during the
afterglow phase, were $1.59\pm0.36$ and $0.98\pm0.17$ respectively.  These
values are significantly larger than the BAT spectral index during the burst
($\beta=0.44$, and $\beta=0.23$ for GRB050126 and GRB050219A respectively)
even when we include the spectral steeping by 1/2 due to electron cooling. If
we assume that toward the end of the burst $\beta$ increased and reached the
value we observe during the afterglow (as appears to be the case for
GRB050219A), then the above argument suggests that the high-angle emission
dominated the observed flux during the adiabatic expansion of shells in the
internal shock, and $\alpha=2+\beta$. Thus, the internal shock model for the
steep X-ray lightcurve has the same characteristic spectral and temporal
features as the model involving high-angle emission in external shocks
described above.

\subsection{Redshift and luminosity}

Adopting the host galaxy redshift $z$ of 1.29 for GRB050126 (Berger \&
Gonzalez 2005) and a BAT fluence of $(1.7\pm0.06)\times 10^{-6}$~erg~cm$^{-2}$
we derive an isotropic gamma-ray energy of
$E_{iso}=1.1\times10^{52}$~erg in the 15-350~keV band (assuming a WMAP
Cosmology of $H_{\rm 0}=70$~km~s$^{-1}$~Mpc$^{-1}$, $\Omega_{\lambda}=0.73$,
and $\Omega_{m}=1-\Omega_{\lambda}$).  For GRB050219A there was no
ground-based optical/IR detection and hence no redshift estimate for this
source. However, we can estimate the redshift (with large uncertainty) using
the relationship $|\Gamma_{\gamma}|= (2.76\pm0.09)(1+z)^{-0.75\pm0.06}$ (Amati
et al 2002; Moran et~al. 2004). For GRB050219A ($\Gamma_{\gamma}=1.23\pm0.06$)
we estimate a redshift of $1.94^{+0.69}_{-0.49}$, which for a BAT fluence of
$(5.2\pm0.4)\times10^{-6}$~erg~cm$^{-2}$ implies an isotropic gamma-ray energy
$E_{\rm iso}=9.5\times10^{52}$~erg in the 15--350~keV band.

Re-writing the Amati relation (Amati et~al. 2002) in terms of the isotropic
energy, $E_{\rm iso}$, and peak energy $E_{\rm peak}$, ie. $E_{\rm
peak}=95(E_{\rm iso}/10^{52}~{\rm erg})^{0.52}$, (Friedman and Bloom 2005), we
derive a rest-frame $E_{\rm peak}> 100$~keV for GRB050126.  Using the
Ghirlanda relation (Ghirlanda, Ghissellini \& Lazzati 2004; Friedman \& Bloom
2005) which describes the dependence of the total energy radiated $E_{\gamma}$ in $\gamma$-rays on $E_{\rm peak}$, where $E_{\rm peak}=512(E_{
\gamma}/10^{51}~{\rm erg})^{0.7}$, we derive an $E_{\gamma}> 9.8\times
10^{49}$~erg. This implies a beaming fraction $f_{\rm B}= E_{\gamma
}/E_{\rm iso}>0.009$.  For GRB050219A, we measure a mean rest-frame peak
energy of $E_{\rm peak}>285.2$~keV using our adopted redshift of 1.94, close
to the predicted rest-frame $E_{\rm peak}$ of 305.6~keV found using the
Ghirlanda relation. Using these estimates for $E_{\rm peak}$ we derive $E_{
\gamma}>4.8\times 10^{50}$~erg, and beaming fraction $f_{\rm B}=0.005$.

As far as these relationships are applicable to GRBs in general\footnote{The
Ghirlanda relation is derived from bursts the majority of which have
$E_{peak}$ outside of the 10-100~keV range, where the BAT has the largest
effective area.}, then both $E_{\rm{iso}}$ and $E_{\gamma}$ for GRB050126
place it amongst the low-end of the $E_{\rm{iso}}$ and $E_{\gamma}$
distributions given in Frail et~al. (2001), while for GRB050219A $E_{\rm iso}$
and $E_{\gamma}$ lie near the mean of these distributions.  We note that the
derived beaming fraction for both sources is a factor of a few larger than the
mean beaming fraction ($f_{\rm B}=0.002$) found by Frail et~al. for the same
sample. The predicted break timescales, 11.2 days and 2.4 days for GRB050126
and GRB050219A respectively, both occur later than the last XRT observation
for these sources and are consistent with the non-detection of a jet-break in
their lightcurves.

\section{Summary and Conclusions}

The early X-ray decay light-curves of GRB050126 and GRB050219A are
characterised by an unusually steep early decline ($f_{\nu}(t)\propto
t^{-3}$), flattening to a more gradual decline on timescales of a few hundred
seconds.

The prompt $\gamma$-ray and the early X-ray afterglow emissions for GRB 050126
\& 050219A require at least two, possibly three, distinct mechanisms. The
X-ray lightcurve after the break for these GRBs is produced in the forward
external shock. The early, steeply falling, X-ray lightcurve may be due to
synchrotron self-Comptonisation in the reverse shock. However, in order to
avoid early bright optical emission from these bursts, which was not seen in
the UVOT data, we require the GRB ejecta to be highly enriched with $e^\pm$
pairs, and to have a Lorentz factor of at least a few hundred.  Furthermore,
it is difficult to see how SSC can account for a steep decay over almost 2
orders of magnitude. An alternate possibility is that the early X-ray
light-curve was produced in the external shock from a jet consisting of narrow
regions (regions of angular size $\le\Gamma^{-1}$) of high energy density.

We suggest that the most plausible explanations for the steep early decay are
either high-latitude ($\theta>\Gamma^{-1}$) emission from a relativistic jet
arriving at the observer when emission from $\theta<\Gamma^{-1}$ has dropped
to zero (the curvature effect) or X-ray flares, indicative of late time
activity of the central engine.


\begin{acknowledgements}

MRG would like to thank the anonymous referee for a prompt and thorough
reading of this manuscript. This work is supported at the University of
Leicester by the Particle Physics and Astronomy Research Council (PPARC), at
Penn State by NASA contract NAS5-00136, and in Italy by funding from ASI on
contract number I/R/039/04.
\end{acknowledgements}

\email{mrg@star.le.ac.uk}.


\clearpage


\begin{table*}
\begin{center}
\caption{Swift BAT spectral fits. Quoted errors are 90\% confidence on 
1 interesting parameter\label{table_bat}}
\begin{tabular}{ccc}
\hline\hline
 & \multicolumn{1}{c}{GRB050126} & \multicolumn{1}{c}{GRB050219a} \\
Parameter &  & \\
 &  & \\
T90 (s)       & $25.7\pm0.1$ & $23.5\pm0.02$ \\
Fluence ($\times10^{-6}$~erg~cm$^{-2}$)  & $1.7\pm0.3$ & $5.2\pm 0.4$ \\
 &  & \\
\hline
& \multicolumn{2}{c}{Model 1: Power-law (20--150keV)} \\
 &  & \\
$\Gamma$  & $1.44\pm{0.18}$ & $1.23^{+0.06}_{-0.06}$ \\
$\chi^{2}/dof$ & 66.2/53 & 125/53 \\
 &  & \\
\hline
& \multicolumn{2}{c}{Model 2: Cut-off powerlaw (20--150~keV)} \\
 &  & \\
$\Gamma$     & -- & $-0.39^{+0.38}_{-0.40}$ \\
$E_{\rm peak}$   & -- & $97.0^{+51.1}_{-31.8}$ \\
$\chi^{2}/dof$ & -- & 46.9/52 \\
 &  & \\
\hline
\end{tabular}
\end{center}
\end{table*}

\begin{table*}
\begin{center}
\caption{Swift XRT temporal decay fits. Quoted errors are 90\% confidence on 1
  interesting parameter.\label{table_decay}}
\begin{tabular}{cccc}
\hline\hline
 & \multicolumn{1}{c}{GRB050126} & \multicolumn{1}{c}{GRB050219a} \\
Parameter & \multicolumn{2}{c}{Model 1: Powerlaw  ($f(t)\propto t^{-\alpha}$)} \\
& & \\
$\alpha$         & $2.52^{+7.5}_{-0.3}$ & $2.50^{+0.16}_{-0.16}$ \\
$\chi^{2}/\nu$ & --  & 225.1/38 \\
$Cash$ $statistic/ndp^{\dagger}$ & 62.0/20  & -- \\
& & \\
\hline
& \multicolumn{2}{c}{Model 2: Broken powerlaw} \\
& & \\
$\alpha_{1}$     & $2.52^{+0.50}_{-0.22}$  & $3.17^{+0.24}_{-0.16}$ \\
$T_{\rm break}$  & $424^{+561}_{-120}$       & $332.1^{+25.8}_{-21.6}$  \\
$\alpha_{2}$     & $1.00^{+0.17}_{-0.26}$  & $0.75^{+0.09}_{-0.07}$ \\
$\chi^{2}/\nu$ & -- & 74.6/36 \\
$Cash$ $statistic/ndp$ & 26.1/20 & -- \\
& & \\ 
\hline
& \multicolumn{2}{c}{Model 3: Offset powerlaw $f(t)\propto (t-t_{\rm a})^{-\alpha}$} \\
& & \\
$\alpha$  & $1.08^{+0.09}_{-0.09}$    & $1.10^{+0.09}_{-0.08}$ \\
$t_{\rm a}$   & $105.1^{+9.1}_{-11.3}$    & $100.7^{+2.8}_{-4.0}$ \\
$\chi^{2}/\nu$ & -- & 114.1/37 \\
$Cash$ $statistic/ndp$   & 31.7/20 & -- \\
& & \\ 
\hline
& \multicolumn{2}{c}{Model 4: Gaussian + powerlaw} \\
& & \\
$t_{\rm g}$ (fixed) & 0.0 & 0.0 \\
$\sigma$ & $89.1^{+17.8}_{-16.1}$ & $77.9^{+6.2}_{-6.0}$ \\
$\alpha$ & $1.11^{+0.12}_{-0.11}$ & $0.81^{+0.09}_{-0.07}$ \\
$\chi^{2}/\nu$ & -- & 89.4/36 \\
$Cash$ $statistic/ndp$ & 26.4/20 & -- \\
& & \\ 
\hline
\end{tabular}
\end{center}
\noindent $\dagger$ ndp number of data points.
\end{table*}

\begin{table*}
\begin{center}
\caption{Swift XRT spectral fits. Quoted errors are 90\% confidence on 1
  interesting parameter.\label{table_xrt}}
\begin{tabular}{ccc}
\hline\hline
 & \multicolumn{1}{c}{GRB050126} & \multicolumn{1}{c}{GRB050219a} \\
& \multicolumn{2}{c}{Model 1: $wa*po^{\dagger}$ - all$^{a}$} \\
& & \\
Mode             & PC only        &     WT+PC+LrPD  \\
gal$N_{\rm H}$      & $5.3\times10^{20}$     &  $8.5\times10^{20}$  \\
$\Gamma$         & $2.26^{+0.26}_{-0.25}$ & $1.90^{+0.17}_{-0.16}$ \\
$N_{\rm excess}$     &  -- & $1.34^{+0.51}_{-0.47}\times10^{21}$ \\
$\chi^{2}_{\nu}$/dof &  8.2/8 & 55/52  \\
& & \\
\hline
& \multicolumn{2}{c}{Model 2: $wa*po^{\dagger}$ - all tied$^{b}$} \\
& & \\
Mode              &  PC only &         WT only \\
gal$N_{\rm H}$       &  $5.3\times10^{20}$   & $8.5\times10^{20}$  \\
$\Gamma$          &  $2.42^{+0.33}_{-0.31}$ & $1.97^{+0.17}_{-0.16}$ \\
$N_{\rm excess}$  &  --                   &
 $1.53^{+0.52}_{-0.49}\times10^{21}$ \\
$\chi^{2}_{\nu}$/dof  &  8.6/6    & 66/49 \\
& & \\
\hline
& \multicolumn{2}{c}{Model 3: $wa*po^{\dagger}$ - untied$^{c}$} \\
& & \\
Mode              & PC only            &  WT only    \\
gal$N_{\rm H}$       &  $5.3\times10^{20}$ & $8.5\times10^{20}$ \\
$\Gamma_{pre}$    &  $2.59^{+0.38}_{-0.35}$ & $1.98^{+0.18}_{-0.16}$ \\
$\Gamma_{post}$   &  $1.72^{+0.65}_{-0.60}$ & $1.89^{+0.26}_{-0.23}$ \\
$N_{\rm excess}$  & --                  & $1.49^{+0.53}_{-0.48}\times10^{21}$ \\
$\chi^{2}_{\nu}$  & 5.0/5              & 66/48 \\
& & \\
\hline
\end{tabular}
\end{center}
\noindent $\dagger$ Here we use the standard XSPEC notation where $wa*po$
means absorbed powerlaw.\newline  
\noindent $a$ all - simultaneous fit to all data.\newline
\noindent $b$ all tied - simultaneous fit to pre- and post-break spectra
with the photon index in each part of the spectrum tied together.\newline
\noindent $c$ untied - simultaneous fit to pre- and post-break spectra
allowing $\Gamma$ and $N_{\rm excess}$ where applicable to freely vary between
the pre- an post-break data.
\end{table*}

\begin{table*}
\begin{center}
\caption{GRB050219a BAT time-resolved spectral fits.\label{table_grb050219a}}
\begin{tabular}{cccccc}
\hline\hline
 & \multicolumn{1}{c}{segment 1} & \multicolumn{1}{c}{segment 2} &
 \multicolumn{1}{c}{segment 3} & \multicolumn{1}{c}{segment 4} &
 \multicolumn{1}{c}{segment 4$+$5} \\
duration (s) & 8 & 6 & 4 & 7 & 14 \\
$\Gamma$ & $-0.59^{+0.59}_{-0.65}$ & $-0.34^{+0.44}_{-0.58}$ & $0.23^{+0.42}_{-0.47}$ & $0.42^{+0.40}_{-0.56}$ & $0.75^{+0.50}_{-0.56}$ \\
$E_{\rm peak}$$^{\dagger}$ & $90.9^{+9.3}_{-8.0}$  & $94.3^{+8.4}_{-7.5}$ & $123.0^{+18.1}_{-14.1}$ & $67.6^{+7.9}_{-6.8}$ & $62.3^{+9.5}_{-7.4}$ \\ 
$\chi^{2}$/dof & $44.2/55$           & $36.4/55$ & $62.5/55$ & $41.7/55$ & $57.7/55$ \\ 
\hline
\end{tabular}
\end{center}
\noindent $\dagger$ The error on $E_{\rm peak}$ is the 90\% confidence limit
for a fixed value of the high energy cut-off $E_{\rm HighEcut}$, where $E_{\rm
peak}=(2-\Gamma)\times E_{\rm HighEcut}$.\newline
\end{table*}

\begin{table*}
\begin{center}
\caption{The relationship between temporal decay index $\alpha$, and spectral
  slopes $\beta$, for internal and external shock models.\label{table_stuff}}
\begin{tabular}{c|cc|cccc}
\hline\hline
& & &  \multicolumn{4}{c}{Forward shock}\\
Source & $\alpha_{post-break}$ & $\beta_{post-break}$ &  \multicolumn{2}{c}{$p=2\beta$} & \multicolumn{2}{c}{$\alpha=(3\beta-1)/2$}\\
\hline
 & & & & & & \\
GRB050126  & $1.14^{+0.08}_{-0.07}$ & $0.72^{+0.65}_{-0.60}$ &
\multicolumn{2}{c}{ $1.44^{+1.2}_{-1.2}$ } &
\multicolumn{2}{c}{ $0.58^{+0.98}_{-0.90}$ } \\
 & & & & & & \\
GRB050219a & $0.75^{+0.09}_{-0.07}$ & $0.89^{+0.26}_{-0.23}$ &
\multicolumn{2}{c}{ $1.78^{+0.5}_{-0.5}$ } &
 \multicolumn{2}{c}{ $0.83^{+0.40}_{-0.34}$ } \\
 & & & & & & \\
\hline
 & & & \multicolumn{2}{c}{External shock} & \multicolumn{2}{c}{Internal/Reverse shock} \\
 & & & \multicolumn{2}{c}{ ($f(\nu,t)\propto
t^{-p}\nu^{-p/2}$) } & \multicolumn{2}{c}{ ($f(\nu,t)\propto
  t^{-2-\beta}\nu^{-\beta}$) } \\
 & $\alpha_{pre-break}$  & $\beta_{pre-break}$ & $p=2\beta$ & $\alpha=p$ & $\beta$ & $\alpha=2+\beta$ \\
\hline
& & & & & &\\
GRB050126  & $2.82^{+0.55}_{-0.44}$ & $1.59^{+0.38}_{-0.35}$ & $3.18^{+0.76}_{-0.70}$ & $3.18^{+0.76}_{-0.70}$ &
$1.59^{+0.38}_{-0.35}$ & $3.59^{+0.38}_{-0.35}$ \\
 & & & & & & \\
GRB050219a &  $3.17^{+0.24}_{-0.16}$ & $0.98^{+0.18}_{-0.16}$ & $1.96^{+0.36}_{-0.32}$ & $1.96^{+0.36}_{-0.32}$ &
$0.98^{+0.18}_{-0.16}$ & $2.98^{+0.18}_{-0.16}$ \\
\hline
\end{tabular}
\end{center}
\end{table*}

\clearpage





\clearpage


\clearpage



\end{document}